\documentclass[PRB,showpacs,preprintnumbers,amsmath,amssymb]{revtex4}
\usepackage{graphicx}% Include figure files
\usepackage{dcolumn}% Align table columns on decimal point
\usepackage{bm}% bold math
\usepackage{wasysym}
\usepackage{color}

\begin{document}

\title{Hierarchy of fillings for FQHE in monolayer and in bilayer graphene: Explanation of $\nu=-\frac{1}{2}$ fractional quantum Hall state in bilayer graphene}

\author{Janusz Jacak}
\email{janusz.jacak@pwr.edu.pl}
\affiliation{Institute of Physics, Wroc{\l}aw University of Technology, Wyb. Wyspia\'nskiego 27, 50-370 Wroc{\l}aw, Poland}
\author{Lucjan Jacak}
\affiliation{Institute of Physics, Wroc{\l}aw University of Technology, Wyb. Wyspia\'nskiego 27, 50-370 Wroc{\l}aw, Poland}

\date{\today}% It is always \today, today,

\begin{abstract}
The commensurability condition is applied to determine the hierarchy of fractional fillings of Landau levels in monolayer and bilayer graphene. The filling rates for FQHE in graphene are found and illustrated in the first three Landau levels. The good agreement with the experimental data is achieved.  The presence of  even denominator filling fractions in the hierarchy for FQHE in bilayer graphene is explained. 
\end{abstract}

\pacs{Valid PACS appear here}% PACS, the Physics and Astronomy
%Use showkeys class option if keyword 

\maketitle

\section{Introduction} 
Recent progress in Hall experiment with graphene reveals many new features in longitudinal and transversal resistivity in Hall configurations characteristic for fractional quantum Hall effect (FQHE) both in suspended graphene scrapings \cite{bil,feldman,feldman2}, and in graphene samples on crystalline substrate of boron nitride \cite{dean2011,amet}. The different structure of Landau levels (LLs) in graphene in comparison to the conventional semiconductor 2DEG is the source of distinct scheme for integer quantum Hall effect  (IQHE) in graphene referred as to 'relativistic' its version \cite{gr2}. The Berry phase induced  shift for chiral carriers in graphene together with the four-fold spin-valley degeneration of LLs result in $\nu=4\left(n+\frac{1}{2}\right)$ series for fillings at which IQHE plateaus occur in subsequent centers of LLs for the case of  monolayer  graphene.  When $SU(4)$ spin-valley symmetry is broken by larger magnetic field, the new features for IQHE appear corresponding to removing of LL subbands degeneration \cite{dur1}. In bilayer graphene the extra degeneration of $n=0$ and $n=1$ LLs shifts plateau positions to ends of LLs also four-fold degenerated in this case except for the eight-fold degenerated LLL. Simultaneously, still more and more features at fractional fillings of LLs are observed related to FQHE, revealing also a specific its character mostly connected with the different structure of subbands of LLs and particle-hole symmetry  due to meeting of the conductivity and valence bands in Dirac points. The new filling fractions are observed in six first subbands of LLs with $n=0$ and $n=1$ in monolayer graphene \cite{dean2011,amet,feldman,feldman2} not repeating the hierarchy of FQHE in conventional 2DEG and  not allowing for explanation upon standard composite fermion model. Especially interesting is an observation of unusual  even denominator fillings for FQHE in bilayer graphene including the most pronounced feature at $\nu=-\frac{1}{2}$ \cite{bil}, taking into account possibility for control of non-Abelian anyons excited from the $-\frac{1}{2}$ state as previously studied for $\nu=\frac{5}{2}$ in traditional semiconductor 2DEG \cite{5/2-1} and important for potential application in topological quantum information processing.

In the present letter we analyze the details of the hierarchy for fractional fillings linked to strongly correlated multiparticle states in graphene using the topological commensurability approach developed earlier for the ordinary 2DEG Hall systems \cite{wu,epl,jac-ll}. In this way we explain the structure of fractional fillings of LL subbands and demonstrate  its evolution with growing number of the LL and the number of its subband. This approach gives the hierarchy of  FQHE and of other  features, including paired states and Hall metal, in agreement with the 
available experimental data for monolayer and bilayer graphene. In particular the explanation of the even denominator filling ratios for bilayer graphene, with the FQHE  state  at  $\nu=-\frac{1}{2}$, has been achieved in a direct manner upon the applied commensurability topological method. 
  
\section{Commensurability condition---the basic formulation}

The concept of commensurability of classical cyclotron  trajectories with interparticle spacing in 2D systems is born in relation to the braid group approach to multiparticle systems in the presence of a magnetic field. If $N$ identical particles are located on the manifold $M$, the collective behavior of such  system can be assigned by the statistics phase shift if one considers position exchange of particle pair \cite{lwitt,wu}. This quantum feature is associated with the one dimensional unitary representation (1DUR) of the full braid group  related to the system. The full braid group is the first homotopy group of the multiparticle configuration space, $\pi_1(\Phi)=\pi_1((M^N-\Delta)/S_N)$, where $M^N$ is the $N$-fold normal product of the manifold $M$, $\Delta $ is the diagonal point set in this product (when coordinates of two or more  particles coincide, subtracted in order to assure conservation of the number of particles), $S_N$ is the permutation group of $N$ elements \cite{birman}. The quotient structure of the configuration space $\Phi=(M^N-\Delta)/S_N)$ is related with the indistinguishability  of quantum identical particles---the property essential for the quantum statistics determination reason. Apart of this quantum prerequisite the full braid group $\pi_1(\Phi)$ is the classical topological object collecting all classes of nonhomotopic trajectory loops in the configuration space $\Phi$, where points which differ only by enumeration of particles are unified. Any details of the dynamics of $N$ particle system  caused by the interaction  and resulting in  special shapes of trajectories are not important here. Only topology of trajectories decides whether one trajectory loop  can be continuously transformed into another one or not. In the latter case such topologically nonequivalent trajectories fall to the distinct classes of the full braid group. Thus, the full braid group does not reflect the dynamics details but rather identifies the topology restrictions imposed on the multiparticle system associated  to some global features of the system including the manifold $M$ type. As the loops from $\pi_1(\Phi)$ describe exchanges of particles (due to identification of  positions of all particles which differ by the enumeration only) the full braid group contains information on quantum statistics of particles, though classical  particles do not have any such statistics. It was proved that the one dimensional unitary representations (1DURs) of the full braid group serve as determinants of the quantum statistics of quantum particles related to the original classical ones \cite{lwitt}. Therefore, if particle classical positions defined by the coordinates of the multiparticle wave function $\Psi (x_1, ..., x_N)$ ($x_i$ is the coordinate of $i$-th particle   on the manifold $M$, i.e., the classical position of the $i$-th particle on $M$) are changing along a selected loop from the $\pi_1(\Phi)$, then this wave function acquires the phase shift $e^{i \alpha}$ given by the 1DUR of this particular braid \cite{imbo}. In this way the statistics of quantum particles can be identified. For the same classical particles the quantumly different particles can be defined as assigned by distinct 1DURs of the  related full braid group. For three dimensional (or of higher dimension) manifolds $M$ the full braid group is simply the permutation group $S_N$, with only two different 1DURs \cite{birman}, 
\begin{equation} 
\sigma_i\rightarrow \left\{ \begin{array}{l} e^{i0}=1,\\
e^{i\pi}=-1,\\
\end{array}\right. 
\end{equation}
where $\sigma_i$ is the generator of $S_N$, i.e., the elementary braid describing exchange of $i$-th and $i+1$-th particles, whereas other particles are left in their positions. These two 1DURs correspond to bosons and fermions, respectively. 
For two-dimensional
manifolds $M$
the braid groups  differ considerably from the permutation group. Unlike the permutation group, for $ M=R^2$, $\pi_1(\Phi)$  is an infinite group with the 1DURs \cite{birman},
\begin{equation}
\sigma_i\rightarrow e^{i\alpha}, \;\alpha\in[0,2\pi),
\end{equation}
corresponding to so-called anyons (including bosons for $\alpha=0$ and fermions for $\alpha =\pi$). Anyons are quantum particles, different than bosons or fermions, which can appear in 2D multiparticle systems, e.g., for particles located on the $R^2$ plane or on the locally 2D manifold, like a sphere (but not on torus \cite{einarsson}, for which 1DURs of related full braid group do not exist).   

Nevertheless, there are also other peculiarities of 2D manifold topology, which were not accounted for in the manner as described above. The special feature of planar multiparticle systems manifests itself in the presence of the perpendicular magnetic  field strong enough that the classical cyclotron radius is shorter in comparison to the interparticle separation on the plane. Because the classical trajectories of free charged particles are defined by cyclotron orbits at the presence of a magnetic field, thus exchanges of neighboring particles in 2D topology are possible only if the size of the cyclotron orbits fits to the separation between particles, as  is illustrated in Fig. \ref{figg1}. Let us emphasize  that the average separation between particles on the plane is rigidly fixed by the Coulomb interaction between particles preventing approaching one particle onto another one in the uniformly distributed $N$ particles. 
The fulfillment of the commensurability between the cyclotron radius and the interparticle spacing   is required for the definition of the generators $\sigma_i$ of the full braid group, where $\sigma_i$ describes exchange of $i$-th and $i+1$-th particles. If the cyclotron orbit is incommensurate with the separation between particles on the plane, then the classical trajectories describing $\sigma_i$ elements of the full braid group are impossible and this braid group cannot be defined for such strong magnetic field. It means that the quantum statistics cannot be defined in this case and any correlated multiparticle quantum state cannot be  organized.
This situation happens at fractional filling of the lowest Landau level (LLL). For magnetic fields larger that this one which corresponds to the completely filled LLL, the classical cyclotron orbits are too short to match neighboring particles and the definition of the generators of  the full braid group is precluded. Nevertheless, at some magic fractional filling ratios of the LLL the correlated states are experimentally observed and refereed as to fractional quantum Hall effect \cite{tsui1982,wlasne1,prange}. In means that possibility of particle exchanges is recovered somehow. In the framework of the composite fermion (CF) approach \cite{jain} the enlargement of cyclotron orbit size is achieved by screening of the external magnetic field by auxiliary field flux quanta attached to particles. This artificial model allows for identification of the  main line of fractional fillings for FQHE by mapping of the fractional state onto integer quantum Hall states in resultant magnetic field  reduced by the average field of  fluxes pinned to CFs. This model does not explain neither the origin of auxiliary field fluxes nor the mechanism of creation of composite particles. 

The manifestation of FQHE can be explain, however, also in the braid group terms \cite{epl}. Though the generators $\sigma_i$ of the full braid group cannot be defined when the cyclotron orbits are shorter than interparticle separation, there exist other braids which in 2D fit to interparticle separations. This exceptional property of planar multiparticle systems possess multilooped cyclotron braids, i.e., $\sigma_i^q$ where $q$--odd integer, and for e.g.,  $q=3$, $\sigma_i^3$ describes the braid for exchange of $i$-th and $i+1$-th particles with an additional loop. Exclusively in 2D the multilooped cyclotron orbits have larger size which can fit to interparticle separation at the magic filling fractions of the LLL, the same ones at which FQHE is observed. The reason of the  enhancement of the size of planar multilooped orbits is linked with constant surface field of  2D orbits despite its multilooped character (in opposition to 3D case when each additional loop adds also a surface portion spanned by this loop).  When the total external field flux is passing through the 2D multilooped orbit it  must be shared between all loops and  their sizes grow. This is illustrated in Fig. \ref{figg2}. 

In Fig. \ref{figg2} (left) the scheme of the cyclotron orbit at magnetic field $B$ is shown as accommodated to the quantum of the  magnetic  field flux, i.e., $BA=\frac{hc}{e}$. This serves here as the definition of the cyclotron orbit $A$ because  this $A$ fits to the interparticle separation, $\frac{ S}{N}$, $S$---the sample area, $N$---the number of particles, in the case of the completely filled LLL. If only single-loop orbits are available, then at, for instance, 3-times larger field, $3B$, the cyclotron orbit accommodated again to the flux quantum is too short in comparison to  the interparticle separation $\frac{S}{N}=A$ (which fitted to the $B$ field orbits but not to $3B$ orbits). This is illustrated in the central  panel of Fig. \ref{figg2}. Nevertheless, if tree-loop orbits are considered, then in flat geometry of 2D space, the external flux $3BA$ must be shared between three loops  with the same surface $A$ (i.e., $BA$ for each loop).  Thus, each loop accommodated to the flux quantum $\frac{hc}{e}$ has  the orbit with the surface $A$ and gives the contribution $BA$ to the  flux, resulting in the total flux $3BA$ per particle, as needed---this is schematically illustrated in Fig. \ref{figg2} (right). The size of $A$ in the right panel is equal to $A$ in the left panel, which means that the  three-loop orbits fit to the interparticle separation equaled to  $A$.

\begin{figure}[ht]
\centering
\scalebox{1.0}{\includegraphics{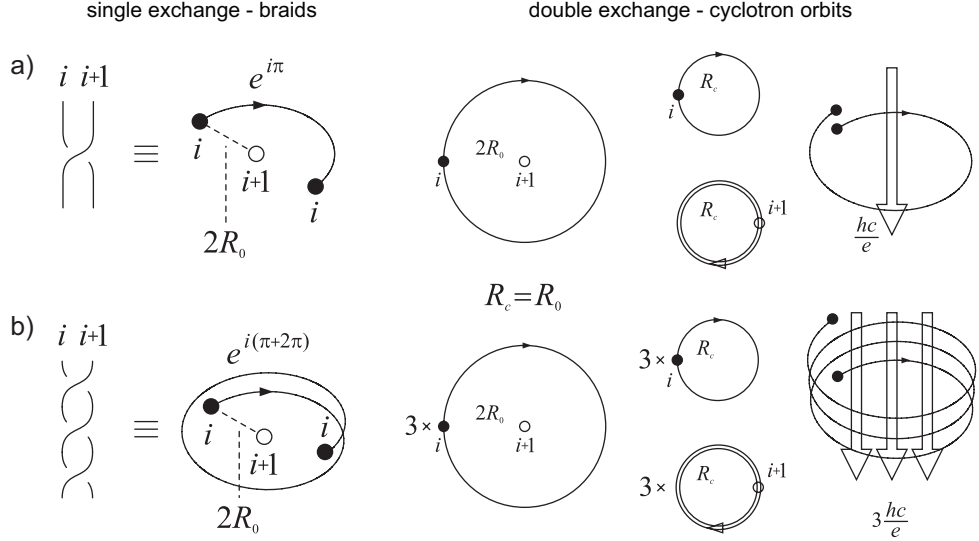}}
\caption{\label{grafenrys5} The braid generator $\sigma_i$ corresponds to the single exchange of particles (left), the cyclotron orbit (relative) corresponds to the double exchange (right), for $\nu=1$ when singlelooped cyclotron trajectory reaches neighboring particles, $R_c=R_o$ (a) braid generator $\sigma_i^3$ for $\nu=\frac{1}{3}$ with additional two loops needed for $R_c=R_o $ ($R_c$--cyclotron radius, $2R_o$--particle separation, i.e., $\pi R_c^2=\frac{hc}{eB}$, $\pi R_o^2=\frac{S}{N}$) (b)} 
\end{figure}

As the braid group generators must be defined by the half of the cyclotron orbit
(cf. Fig. \ref{grafenrys5}) thus the braid with one additional loop corresponds to the cyclotron orbits with three loops---such a generator has the form $b^{(3)}_i=(\sigma_i)^3$ and the braid group generated by $b^{(3)}_i$, $i=1,\dots,N$  is the subgroup of the original full braid group. This subgroup is called the cyclotron braid subgroup and its 1DURs define statistics of 2D charged particles  at strong magnetic field corresponding to fractional filling $\nu=\frac{1}{3}$ of the LLL giving rise to the  explanation of the Laughlin statistics for FQHE at this fractional filling. The generalization to more loops attached to the braid generator one by one, results in double increase of loop number in multilooped cyclotron orbits and thereby in fractions $\nu=\frac{1}{p}$, $p-odd\; integer$. This approach successfully reproduces the hierarchy of the experimentally observed filling fraction corresponding to FQHE in the LLL and in the higher LLs for 2DEG in  conventional semiconductors \cite{jac-ll}.

If  $\frac{S}{N}<A$, i.e., when cyclotron orbits are larger that interparticle separation, (as in the  right panel in Fig. \ref{figg1}), some special commensurability  circumstances important for braid group definition also occur. For the filling  fractions when  $x\frac{S}{N}=A$; $x-integer$, the cyclotron orbits fit  to every  $x$-th particle separation ($x\frac{S}{N}=\frac{S}{N/x}$), which also allows for the  definition of the generators $\sigma_i$ in the form of  ordinary singlelooped braids, similarly as happens for the  completely filled higher LLs. The related statistics is the same as for IQHE though at some fractional fillings of higher LLs as is demonstrated in Ref. \cite{jac-ll} in good correspondence with experimental observations up to the three first LLs  for 2DEG in conventional semiconductors \cite{lls,ll8/3-1,5/2-1}. This could happen only in such LL subbbands where the  condition $x\frac{S}{N}=A$ ( $x-integer$) could be fulfilled. In the conventional semiconductor Hall systems it may happen only  for $n\geq 1$ where $n$ is the number enumerating LLs. Simultaneously, for   $n>0$ too short cyclotron orbits can be encountered only close to subbands borders, because in the higher LLs the  cyclotron singlelooped orbits are larger as accommodated to higher kinetic energy. This pushes FQHE(multiloop) features in higher LLs toward the edges of subbands in  LLs with $n>0$, whereas in central regions of these subbbands the new fractional features occur related with the singleooped cyclotron orbits and thus with  IQHE-type of correlations but at fractional filling ratios. These correlated states are referred as to FQHE(singleloop). The quantization of the  transverse resistance  $R_{xy}$ related to these  fractional filling ratios of higher LLs $\nu=\frac{hc}{eB}$ is as for FQHE $\frac{h}{e^2\nu}$, but the correlations of Laughlin type are with the exponent $p=1$  displaying singlelooped braid exchanges like in IQHE. The number of these new fractional  filling ratios grows as $2n$ with the LL number.

\section{Commensurability induced hierarchy of fractional fillings for graphene}

In graphene one deals with the relativistic version of LLs \cite{geb,gr2}. This is due to specific band structure in this material being a gapless semiconductor with points $K$ and $K'$ on the border of the  hexagonal Brillouin zone where the valence and conduction bands meet together \cite{gr2}. Thus the low energy particle-hole excitations can be described by the effective Dirac Hamiltonian corresponding to cone-shape of both bands close to the meeting points. The quantization due presence of the magnetic field has the form of degenerated LLs, though with spectrum  not equidistant as for ordinary 2DEG but enumerated by $\sqrt{n}$, $n$ is the number of the LL \cite{geb}. This form of the selfenergies results from the linear in momentum Hamiltonian close to $K$ and $K'$ points, whereas the degeneration of each LL subband is the same as for 2DEG and equals to $\frac{BS}{hc/e}$. The number of subbbands per each LL (each $n$) is here 4. This corresponds  to the ordinary Zeeman spin-splitting and to the so called valley splitting   (expressed often in terms of a pseudospin)  due to doublet of inequivalent $K$ points mixed with two sublattices for $C$ atoms in crystal net of graphene \cite{geb}. Taking into account that the Zeeman splitting in graphene is small \cite{gr2} and the valley splitting  depending on the external magnetic field is small as well, the 4-fold  approximate spin-valley additional  degeneration
is employed to determine the filling fractions for relativistic IQHE in graphene,
in the following form $\nu=4 (n+\frac{1}{2})$ in pretty good correspondence with the experimental observations \cite{gr2}. The remarkable difference between this filling rate formula and that one for the ordinary 2DEG (with factor 2 instead of 4, due to only approximate spin degeneration) is the presence of the overall shift by the factor 2 resulting from $\frac{1}{2}$ in the above formula. This shift is due to the so-called Berry phase manifesting in the LLL for graphene and resulting  in  the sharing of states from the LLL between particles and holes from the conduction and valence band, respectively, at the zero energy level. Due to this feature the bottom of the LLL is shifted by 2 (in terms of  the filling factor) upward if counting only negatively charged carriers and oppositely---downward  for   positive holes. The corresponding filling rates for holes from the valence band  are negative mirror refection of those for electrons from the conduction band. Note that changing between particles and holes can be easily achieved in graphene by the shift of the Fermi level around the Dirac point   by application of  a lateral voltage.   

The Berry phase contribution is referred sometimes as to the additional $\pi$ phase shift due to the chirality  induced by the valley pseudospin  when one adiabatically traverses with a selected particle a closed   loop (e.g., along  the semiclassical cyclotron loop) when the momentum  made the $2\pi$ convolution. 

Similar analysis of the LLs can be done for the bilayer graphene \cite{falko,geb}. Due to off-diagonal interlayer hopping term the local Hamiltonian attains again the quadratic form with respect to the momentum. Thus the LL spectrum in bilayer graphene resembles that one for the  ordinary 2DEG with four subbands for each level except for the LLL which is eight-fold degenerated. This extra  degeneration of the LLL arose from the action of the square of an annihilation operator on oscillator states with $n=0$ and $n=1$ \cite{falko,geb}. Due to the sharing of the     LLL between particles and holes, the bottom for uniformly charged carriers is located in the center of the 8-fold quasi-degenerated LLL. This property is associated again with the Berry phase shift for chiral particles, though in the case of bilayer graphene it gives additional $2\pi$ phase shift. Therefore, steps of the relativistic IQHE are located in bilayer graphene again at integer filling rates  of the subsequent LLs whereas for monolayer  graphene were located at half-fillings of the LLs \cite{falko,geb}. 
  
\subsection{Monolayer graphene}
For magnetic field strong enough that $\nu\in (0,1)$ one deals with completely filled valence band hole states of the LLL  and with the fractional filled  subsequent particle states from the conduction band subband $n=0,2\uparrow$ ($ 2$ indicates the  valley pseudospin orientation and $ \uparrow $ indicates  orientation of the ordinary spin). The degeneration of each subband is $N_0=\frac{BS}{hc/e}$ and we have less than $N_0$ of  electrons, $N<N_0$ for fractional fillings. 

The cyclotron orbits in the LLL must be accommodated to the bare kinetic energy $T = \hbar {\omega }_{c} (n+\frac{1}{2} )$
where ${\omega }_{c}=\frac{eB}{mc} $.
The cyclotron orbits have the same size for all particles due to the flat band condition quenching the kinetic energy competition and resulting in the same averaged velocity and the same cyclotron orbit size for all particles at a perpendicular magnetic field presence (though quantumly, a velocity is not well defined since its coordinates do not commute). The cyclotron orbits restrict the topology of all trajectories uniformly in 2D, thus restrict the braid group structure despite of particularities of interaction and other fields including the  crystal field. Therefore for graphene the cyclotron orbit structure is governed by ordinary (the same as for 2DEG)
 Landau levels restrictions even though  a specific band structure with Dirac points being the result of the crystal field highly modifies LLs but not in terms of the bare  kinetic energy. The particularities of graphene quantum dynamics induced by the electric-type interaction  are included in the ordinary part of the Feynman path integral, whereas  the additional summation over topologically nonequivalent trajectory classes concerns the braid group structure the same as for 2DEG upon the magnetic field. The difference between the conventional 2DEG systems and graphene will be  related in this regard  with distinct number of  LL subbands  in graphene comparing to the conventional semiconductor  2DEG and in the Barry phase induced shift of fillings in the LLL. 
 
Thus, the cyclotron orbit size in the subband $n=0,2\uparrow$ equals to,
$\frac{hc/e}{B}= \frac{S}{N_0}$, where $S$ is the sample surface. As this orbit size is lower than the interparticle separation $\frac{S}{N}$ (as $N<N_0$), the multilooped braid structure is necessary. From the commensurability condition $ q \frac{S}{N_0}=\frac{S}{N}$ one finds $\nu =\frac{N}{N_0}= \frac{1}{q}$, 
($q-odd\;integer$ to protect the braid structure). For holes in this subband one can expect the symmetric filling ratios $\nu = 1-\frac{1}{q}$. Similarly as for the ordinary 2DEG one can generalize this simple series by assumption that the last loop of the multilooped cyclotron orbit can be commensurate with the interparticle separation as for some other filling ratio expressed by  $l$, whereas the former loops take away  an integer  number of flux quanta. In this way one can obtain the hierarchy of fillings for FQHE in this subband of the  LLL,
$\nu= \frac{l}{l(q-1)\pm 1},\;\; \nu=1- \frac{l}{l(q-1)\pm 1}$, where $l= 1,2,\dots$ or equals to other fractional filling factor and minus in the denominators indicates possibility of the eight-figure orientation  of the last loop with respect to the antecedent  one. The Hall metal states can be characterized by the limit $\l\rightarrow \infty$ in the above formula  (i.e., for zeroth flux taken away by the last loop
as for ordinary fermions without the magnetic field, which was the case in the Hall metal archetype for $\nu=\frac{1}{2}$ in the conventional 2DEG), which gives the hierarchy for the Hall metal   states, $\nu=\frac{1}{q-1},\;\;\nu=1-\frac{1}{q-1}$.
To account for the Berry phase anomaly in graphene the overall shift of $\nu$ by $-2$ can be  performed, but we use here the net filling fractions. 

For the  completely filled subband $n=0,2\uparrow$, i.e., for $\nu=1$, one arrives at IQHE. For lower magnetic field (or larger number of electrons), when three first subband of the LLL are filled and in the last subband of LLL, the cyclotron orbit size $\frac{S}{N_0}$ is still lower than the interparticle separation $\frac{S}{N-N_0}$ (because $N-N_0<N_0$). Thus multilooped structure is repeated from  the previous subband. This results with the same FQHE hierarchy as for antecedent subband, only shifted ahead by 1. For the completely filled LLL (i.e, for completely filled all its subbands) one deals with IQHE according to its main-line $\nu=4(n+\frac{1}{2})$.

Similarly one can consider fillings of the second LL with $n=1$. This level also has four subbands, but in this level the bare kinetic energy is equaled to $\frac{3\hbar\omega_B}{2}$ and the related cyclotron orbit size $\frac{3hc}{eB}=\frac{3S}{N_0}$. For  $N\in(2N_0,3N_0]$ we deal with graduate filling of $n=1,1\uparrow$ subband. Cyclotron orbits of size $\frac{3 S}{N_0}$
are comparing here to interparticle separation scale $\frac{S}{N-2N_0}$. Only for small number of electrons in this subband one deals with the multilooped structure (corresponding to the inequality $\frac{3S}{N_0}<\frac{S}{N-2N_0}$), when
$q \frac{3 S}{N_0}= \frac{S}{N-2N_0},\;q-odd \;integer$, which gives main series for FQHE(multiloop) in this subband, $\nu= 2+\frac{1}{3 q}$. Similarly as before the complete related hierarchy reads, $\nu= 2+\frac{l}{3l (q-1) \pm 1},\;\;
\nu= 4-\frac{l}{3l (q-1) \pm 1}$, with the  Hall metal hierarchy in the limit $l\rightarrow \infty$. These series are located closer to the subband edges, whereas in the center of this subband the other commensurability conditions are  possible. 
When $ \frac{3 S}{N_0}= \frac{x S}{N-2N_0}$ and $x=1,2,3$ we get $\nu=\frac{7}{3}, \frac{8}{3}, 3$, respectively, corresponding to singlelooped cyclotron orbits similar as for IQHE. Thus, for $\nu=\frac{7}{3}, \frac{8}{3}$ one deals with FQHE(singleloop)---the new Hall feature manifesting itself only in higher LLs, where cyclotron orbits may be larger than the interparticle separation.  At the special case  $ \frac{3 S}{N_0}= \frac{1.5  S}{N-2N_0}$ one can arrive at $\nu=\frac{5}{2}$ with the paired particles (pairing does not change the cyclotron radius but twice diminishes the carrier number to $\frac{N-2N_0}{2}$). 

The following subband  are filled with electrons upon the similar scheme. For the subband $n=1,1\downarrow$ the cyclotron size is $\frac{3S}{N_0}$, whereas the interparticle distances are measured with the plaque $\frac{S}{N-3N_0}$, where $N\in(3N_0,4N_0)$. The commensurability condition 
$\frac{q3S}{N_0}=\frac{S}{N-3N_0}$ results in the main series for FQHE(multiloop), $\nu=3+ \frac{1}{3q}$, which can be developed to the full hierarchy similarly as described above.  The condition $\frac{3S}{N_0}=\frac{xS}{N-3N_0}$ with $x=1,2,3$ results in fractions with singleloop correlations of FQHE(singleloop)-type for  $\nu= \frac{10}{3}, \frac{11}{3}$ and IQHE for $\nu= 4$, correspondingly,  whereas a paired state can be realized at $\nu=\frac{7}{2}$.

\subsection{Bilayer graphene}

\begin{figure}[ht]
\centering
\scalebox{1.0}{\includegraphics{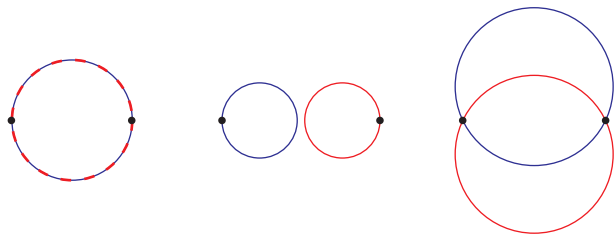}}
\caption{Schematic demonstration that commensurability (left) of cyclotron orbit with interparticle separation satisfies topology requirements for braid interchanges in equidistantly uniformly distributed 2D particles; for smaller cyclotron radii particles cannot be matched (center), for larger ones the interparticle distance cannot be conserved (right) }
\label{figg1}
\end{figure}

The special topology of the bilayer 2DEG structure creates opportunity to verify the braid group based concept of the commensurability of cyclotron orbits size with the interparticle spacing in planar system of interacting particles. The interaction preserves particles do not approach  one to another one closer than particle separation resulting from the planar density.  Only when the cyclotron orbits fit accurately to this spacing the mutual classical exchange of neighboring particles is possible at presence of the perpendicular magnetic field. This is caused by the fact that the trajectories of particles are governed by the cyclotron orbital action of the magnetic field and despite of the interaction particularities the geometry of free particle cyclotron orbits allow for topological classification of possible or not possible  trajectories creating the fundamental $\pi_1$ group of the configuration space for the system of N particles on the plane, i.e., the full braid group for the plane \cite{birman}. In 2D all cyclotron orbits are planar and in the case of uniformly distributed particles with the same velocities exposed to the perpendicular magnetic field they either admit topologically exchanges of neighboring particles   or not depending on the commensurability of the cyclotron orbits with the interparticle separation. If the planar size of the cyclotron orbits $A$ equals to the plane fraction per single particle  $A=\frac{S}{N}$ (where $S$ is the surface field of the sample and $N$ is the number of particles), the mutual exchanges of particles are possible and the ordinary full braid group $\pi_1(\Phi)$ can be defined.  Otherwise, when $\frac{S}{N}>A$, the exchanges of particles are impossible along ordinary cyclotron trajectories, as they are too short to merge neighbors on the plane. When $\frac{S}{N}<A$ the interchanges of 2D particles are also impossible along cyclotron trajectories, because such an exchange do not allow for conservation of uniform particle distribution with the constant interparticle spacing. These three distinct commensurability situations were depicted in Fig. \ref{figg1}. Topological conditioning of  interparticle exchanges in the uniformly distributed  planar systems of charged particles plays the fundamental role in discrimination of trajectories from the full braid  group which are not admitted at the presence  of the magnetic field as do not merge particles. Miss mash between size of cyclotron orbits and interparticle spacing precludes the definition of the full braid group generators $\sigma_i$ corresponding to the interchange of $i$-th and $i+1$-th particles (upon a selected certain  enumeration of particles, arbitrary in general due to particle indistinguishability). When the braid group cannot be defined the collective state cannot be created as the statistics of particles cannot be determined. In order  to establish  the statistics of quantum particles, the related braid group must be defined and the statistics is governed by the selected 1DUR of this group \cite{lwitt}.

\begin{figure}[ht]
\centering
\scalebox{1.0}{\includegraphics{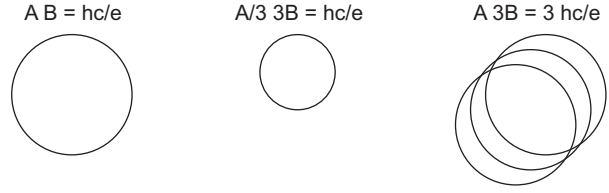}}
\caption{Schematic illustration of cyclotron orbit enhancement in 2D due to multi-loop trajectory structure (third dimension added for visual clarity) }
\label{figg2}
\end{figure}

For the left panel case in Fig. \ref{figg1} the full braid group is well defined and all statistics typical for 2D are available, i.e., according to the form of 1DURs for the $\pi_1(\Phi),\;M=R^2$, $e^{i \alpha}$ with $\alpha\in [0,2\pi)$ corresponding to fermions for $\alpha=\pi$, to  bosons for $\alpha =0$ and to anyons for other $\alpha$. If, however,  $\frac{S}{N}>A$ (as in the  central panel in Fig. \ref{figg1}),  the ordinary  generators $\sigma_i$  cannot be defined. If one removes from the full braid group all  impossible singlelooped-related trajectories corresponding to generators of this group, which cannot be defined as the cyclotron orbits are too small to match particles, one can observe  that the remaining  braids with additional loops can match particles on 2D separated  larger than  the singlelooped cyclotron orbit reaches. These braids are related to multilooped cyclotron orbits with $p$ loops, i.e., when  $pA=\frac{S}{N}$, with $p$ odd integer \cite{epl}. This property follows from the fact that multilooped cyclotron orbits on the plane must be larger than the singlelooped orbits for the same magnetic field. The external magnetic field flux passing  through the multilooped 2D cyclotron orbit is the same as that one passing through  the singlelooped orbit and in the former case, per each loop falls only fraction of the total flux. It means that the size of each loop grows to fit to the fraction of the flux as illustrated in Fig. \ref{figg2}. Repeating again the related argumentation we show the cyclotron orbit  at magnetic field $B$
in Fig. \ref{figg2} (left)  as accommodated to the quantum of the  magnetic  field flux, i.e., $BA=\frac{hc}{e}$, because  the cyclotron orbit $A$ fits to $\frac{ S}{N}$, $S$---the sample area, $N$---the number of particles,  in the case of the completely filled LLL. At $ p$-times larger field, $pB$, the cyclotron orbit accommodated again to the flux quantum is too short in comparison to the interparticle separation $\frac{S}{N}$, as is illustrated in the central  panel of Fig. \ref{figg2} for $p=3$. If $p$-looped orbits are considered, then in of 2D space, the external flux $pBA$ must be shared between $p$ loops  with the same surface $A$ (i.e., $BA$ for each loop).  Thus, each loop accommodated to the flux quantum $\frac{hc}{e}$ has  the orbit with the surface $A$, resulting in the total flux $pBA$ per particle, as is  illustrated in Fig. \ref{figg2} (right). The size of $A$ in the right panel is equal to $A$ in the left panel, which means that the  $p$-looped orbits fit to the interparticle separation defined by $A$ though the singlelooped not. This property attributed exclusively to the  exact 2D topology allows for explanation of FQHE and related exotic Laughlin correlations \cite{epl}.

The bilayer graphene is, however, not strictly two dimensional and 
for the bilayer graphene the topological situation changes considerably. Two sheets of the graphene plane lie in close distance with the hopping constant allowing changing of electron positions between the planes. Thus, we deal   here with double number of electrons residing on two-sheet structure instead of the single sheet as was  the case for the monolayer graphene. 

All above described requirements to fulfill commensurability condition in order to define the related braid group describing the correlated multiparticle state are in charge also for the bilayer graphene, with a single distinction with respect to the monolayer case. The doublelooped cyclotron orbits may have in bilayer graphene the same size as the singlelooped orbit. This follows from the fact that the second loop can be located in the opposite sheet of graphene than the first one and the external field passing through such doublelooped orbit is twice larger than the flux passing through the singlelooped orbit.  Each loop has in this case the separate own surface in contrary to the  multilooped cyclotron orbit  located   on the purely 2D plane which forced   each loop to take   away only  some fraction of the total flux because all these loops share the same surface in 2D. Taking into account that in the bilayer system  loops of the multilooped orbit may be located partly in both 2D sheets, the contribution of the one loop must be avoided whereas the remaining loops must share the same flux as passing through a singlelooped orbit, independently how loops are distributed between two sheets.  Thus, one can write out  the commensurability condition in the bilayer graphene  for the case of too short singlelooped  cyclotron orbits in the following form (for concreteness in the subband $n=0,2\uparrow$ of the LLL---the first particle-type subband of the LLL),
\begin{equation}
\label{21}
\begin{array}{l}
\frac{hc}{eB}=\frac{S}{N_0}<\frac{S}{N-N_0},\\
(p-1)\frac{hc}{eB}=(p-1)\frac{2S}{N_0}=\frac{S}{N-N_0},\\
\nu=\frac{N}{N_0}=\frac{1}{p-1}=\frac{1}{2},\frac{1}{4},\frac{1}{6}, \dots,\\
\end{array}
\end{equation}
where, $N$ is the total number of particles  in both graphene sheets, $N_0$ is the degeneration counted for both sheets together, $S$ is the surface of the sample (the surface of the single sheet)  $p$ is an odd integer to assure that the half of the cyclotron orbit defines the braid.

The factor $p-1$ in l.h.s. of Eq. (\ref{21}) is caused by the fact  that in  enlarging of the effective cyclotron orbits participate only orbits from the ideally 2D sheet of bilayer graphene (no matter in which are located doubling loops) with the exception of a single orbit which is located in opposite sheet to the first one. This sole loop contributes to the total flux with the additional flux quantum due to its own surface and this loop  must be omitted. The next orbits must duplicate the former ones (in fact two) without rising the surface and no matter in which sheet are they located, because in both they will duplicate loops already present there. Thus, in the enhancement of the effective $p$-looped cyclotron orbit take part  only $p-1$ loops.

\begin{figure}[ht!]
\centering
\scalebox{0.6}{\includegraphics{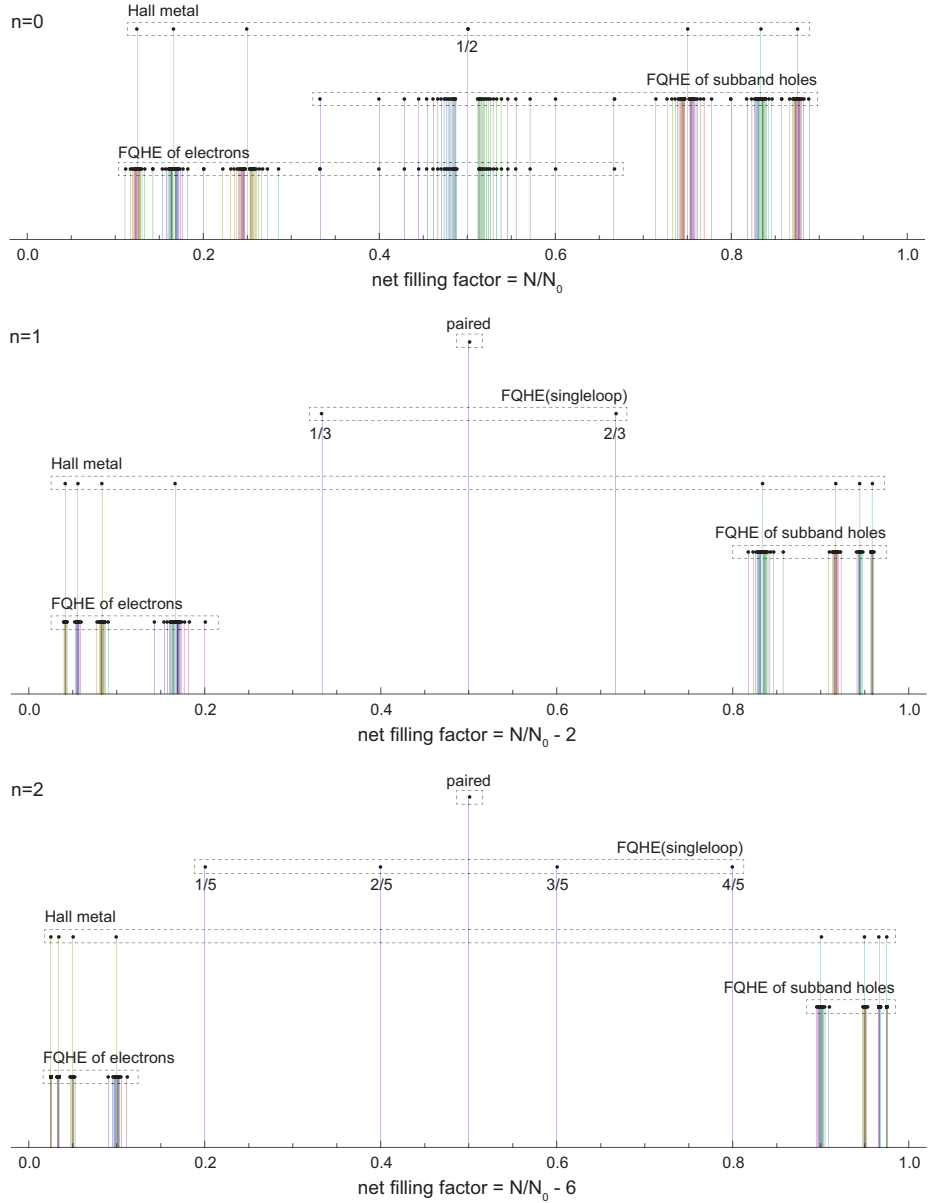}}
\caption{\label{figg4} Evolution of fractional filling hierarchy in three first LLs of monolayer graphene; for each LL the first particle subband is illustrated, the next subbands in each LL repeat the hierarchy from the first one. Different types of ordering are indicated with spikes of various height. Series for FQHE (ordinary---multiloop), FQHE(singleloop), Hall metal and paired state  are displayed acc. to the hierarchy described in Tab. \ref{tab1} with $q=3-9$, $l=1-20$; only a few selected ratios from these series  are explicitly  written out.   }
\end{figure}

Let us emphasize that for such multilooped orbits the total  number of loops  still  is $p$---thus the generators of the corresponding cyclotron subgroup are of the form $b^{(p)}_i=\sigma_i^p$, resulting in the Laughlin correlations with the $p$ exponent for the Jastrow polynomial. But due to the distinct commensurability of orbits with interparticle separation (\ref{21}) the related filling fraction are $\nu=\frac{1}{p-1}$ in the first particle-type subband of the LLL, i.e., the subband $n=0,1,\uparrow$. This even denominator  main series of the FQHE hierarchy for bilayer graphene pretty well coincides with the experimental observations \cite{bil}.

For holes in this subband (let us emphasize that these holes are not holes from  the  valence band but correspond to  unfilled states in the almost filled subband of particle type) one can write,
$\nu = 1-\frac{1}{p-1}$, whereas the generalization to the full hierarchy of FQHE in this subband attains the form, 
$\nu = \frac{l}{l(p-2)\pm 1},\; \;\nu = 1-\frac{l}{l(p-2)\pm 1}$, where $l$ corresponds to some filling factor for other correlated Hall state, including completely filled LLs with IQHE. 
In the next subbands of the LLL, $n=0,2\downarrow$ (assuming tht this subband succeeds the former one), the hierarchy is  identical only shifted  by one, because commensurability condition has the same form for all  subbands with the same $n$  due to  the same size of the cyclotron orbits.

Some novelty occurs, however, in the following two subbands of the LLL, $n=1,2\uparrow$ and $n=1,2\downarrow$. The FQHE main series in the first of these  subbands of the LLL, $n=1,2\uparrow$, has the form,
\begin{equation}
\label{22}
\begin{array}{l}
\frac{3hc}{eB}=\frac{3S}{N_0}<\frac{S}{N-2N_0},\\
(p-1)3\frac{hc}{eB}=(p-1)\frac{3S}{N_0}=\frac{S}{N-2N_0},\\
\nu=\frac{N}{N_0}=2+\frac{1}{3(p-1)}=2+\frac{1}{6},2 +\frac{1}{12},4+ \frac{1}{18}, \dots,\\
\end{array}
\end{equation}
The generalization of this main series for holes in the subband and to the full FQHE hierarchy in this subband is as follows: 
for subband holes (i.e., for not completely filled subband),
$\nu=3-\frac{1}{3(p-1)}$ and for the full FQHE hierarchy in this subband,
$\nu=2+\frac{l}{3 l(p-2)\pm 1},\;\; \nu=3-\frac{l}{3l(p-2)\pm 1}$ (with Hall metal hierarchy in the limit $l\rightarrow \infty$).

\begin{table}[h!]
\centering
\begin{tabular}{|p{1.6cm}|p{5.6cm}|p{6.5cm}|p{3.5cm}|}
\hline
LL  subb.& FQHE(singleloop), paired, IQHE&FQHE(multiloop) ($q-odd,\; l=1,2,3,\dots)$
&Hall metal \\
\hline
$n=0,2\uparrow$&$\nu=$ 1 & 
$\nu=\frac{1}{q}, 1-\frac{1}{q},\frac{l}{l(q-1)\pm1},1-\frac{l}{l(q-1)\pm 1}$&$\nu=\frac{1}{q-1},
1-\frac{1}{q-1}$ \\
\hline
$n=1,1\uparrow$&$\nu=\frac{7}{3},\frac{8}{3}$,
($\frac{5}{2}\; paired$), 2,3& $\nu=2+\frac{1}{3q}$,
$3-\frac{1}{3q}, 2+\frac{l}{3l(q-1)\pm 1}, 3-\frac{l}{3l(q-1)\pm 1}$& $\nu=2+\frac{1}{3(q-1)},
3-\frac{1}{3(q-1)} $
\\
\hline
$n=2,1\uparrow$&$\nu=\frac{31}{5},\frac{32}{5}, \frac{33}{5},\frac{34}{5}$,
($\frac{13}{2}\; paired$), 6,7& $\nu=6+\frac{1}{5q},6+\frac{l}{5l(q-1)\pm 1},7-\frac{1}{5q}, 7-\frac{l}{5l(q-1)\pm 1} $& $\nu = 6+\frac{1}{5(q-1)},7-\frac{1}{5(q-1)}$ \\
\hline
\end{tabular}
\caption{LL filling factors for FQHE determined by commensurability condition ($paired$ indicates condensate of electron pairs), for the  first particle subband in each of the three first LLs ($n=0,1,2$) for monolayer graphene}
\label{tab1} 
\end{table}

\begin{table}[h!]
\centering
\begin{tabular}{|p{2cm}|p{4.7cm}|p{7.4cm}|p{3.2cm}|}
\hline
LL  subb.& FQHE(singleloop), paired, IQHE, $\nu=$&FQHE(multiloop) ($q-odd,\; l=1,2,3,\dots)$, $\nu=$
&Hall metal, $\nu=$ \\
\hline
$n=0,2\uparrow$ (0)& 0,1 & 
$\frac{1}{(q-1)}, 1-\frac{1}{(q-1)},\frac{l}{l(q-2)\pm 1},1-\frac{l}{l(q-2)\pm 1}$&$\frac{1}{q-2},
1-\frac{1}{q-2}$ \\
\hline
$n=1,2\uparrow$ (0)&$\frac{7}{3},\frac{8}{3}$,
($\frac{5}{2}\; paired$), 2,3& $2+\frac{1}{3(q-1)}$,
$3-\frac{1}{3(q-1)}, 2+\frac{l}{3l(q-2)\pm 1}, 3-\frac{l}{3l(q-2)\pm 1}$& $2+\frac{1}{3(q-2)},
3-\frac{1}{3(q-2)} $
\\
\hline
$n=2,1\uparrow$ (1)&$\frac{21}{5},\frac{22}{5}, \frac{23}{5},\frac{24}{5}$,
($\frac{9}{2}\; paired$), 4,5& $4+\frac{1}{5(q-1)},4+\frac{l}{5l(q-2)\pm 1},5-\frac{1}{5(q-1)}, 5-\frac{l}{5l(q-2)\pm 1} $& $4+\frac{1}{5(q-2)},5-\frac{1}{5(q-2)}$ \\
\hline
\end{tabular}
\caption{LL filling factors for  FQHE determined by commensurability condition ($paired$ indicates condensate of electron pairs), for the  first particle subband in each of the two first LLs ($n=0,1$ for the extra degenerated LLL and $n=2$ for the first LL above the LLL) for bilayer graphene}
\label{tab2} 
\end{table}

\begin{table}[h!]
\centering
\begin{tabular}{|p{2cm}|p{4.7cm}|p{7.4cm}|p{3.2cm}|}
\hline
LL  subb.& FQHE(singleloop), paired, IQHE, $\nu=$&FQHE(multiloop) ($q-odd,\; l=1,2,3,\dots)$, $\nu=$
&Hall metal, $\nu=$ \\
\hline
$n=0,2\uparrow$ & 1 & 
$\frac{1}{(q-1)}, 1-\frac{1}{(q-1)},\frac{l}{l(q-2)\pm 1},1-\frac{l}{l(q-2)\pm 1}$&$\frac{1}{q-2},
1-\frac{1}{q-2}$ \\
\hline
$n=1,2\uparrow$&$\frac{4}{3},\frac{5}{3}$,
($\frac{3}{2}\; paired$), 1,2& $1+\frac{1}{3(q-1)}$,
$2-\frac{1}{3(q-1)}, 1+\frac{l}{3l(q-2)\pm 1}, 2-\frac{l}{3l(q-2)\pm 1}$& $1+\frac{1}{3(q-2)},
2-\frac{1}{3(q-2)} $
\\
\hline
\hline
$n=1,2\uparrow$ &$\frac{1}{3},\frac{2}{3}$, 
($\frac{1}{2}\; paired$), 1& $\frac{1}{3(q-1)},\frac{l}{3l(q-2)\pm 1},1-\frac{1}{3(q-1)}, 1-\frac{l}{3l(q-2)\pm 1} $& $\frac{1}{3(q-2)},1-\frac{1}{3(q-2)}$ \\
\hline
$n=0,2\uparrow$& 1,2&$1+\frac{1}{q-1},1+\frac{l}{l(q-1)\pm 1},2-\frac{1}{q-2},2-\frac{l}{l(q-2)\pm 1}$&$1+\frac{1}{q-2},2-\frac{1}{q-2}$\\
\hline
\end{tabular}
\caption{Comparison of filling hierarchy in the LLL level in bilayer graphene for two mutually inverted  successions of two lowest subbands: $n=0,2\uparrow$, $n=1,2,\uparrow$ (upper) and 
$n=1,2\uparrow$, $n=0,2\uparrow$ (lower)
}
\label{tab3} 
\end{table}

Nevertheless, in the subband $n=1,2\uparrow$ of the  LLL some new commensurability opportunity occurs:
$ \frac{3}{N_0}=\frac{x}{N-2N_0}$ for $x=1,2,3$, which gives fillings ratios 
$\nu=\frac{7}{3}, \frac{8}{3}, 3$, correspondingly. All this ratios are related with singlelooped cyclotron trajectories, thus with singleloop correlations similar as for IQHE (though for not integer filling ratios). This new Hall feature, typical for LLs with $n\geq 1$, we called as FQHE(singleloop).  Moreover, for $x=1.5$ one can consider twice diminishing of particle number 
$(N-2N_0)/2$ due to the pairing, which gives perfect commensurability of cyclotron orbits of pairs with the  separation of the particle pairs  at $\nu=\frac{5}{2}$. 

The last  subband $n=1,2\downarrow$ in the LLL in bilayer graphene is filled with electrons  in the similar manner because  for both subbands with $n=1$ the cyclotron orbits have the same size. Thus the hierarchy of fractional filling for the last subband in the LLL is shifted by 1 from the antecedent subband without any modification. The situation changes, however, in the next LL (the first one above the LLL). In the first such LL (with $n=2$) the cyclotron orbits suited to commensurability condition are determined by the bare kinetic energy for $n=2$, and the corresponding cyclotron orbit size is equaled to, $\frac{5 hc}{eB}=\frac{5S}{N_0}$. The similar analysis as in the previous LL, gives here the main series and the full hierarchy for FQHE(multiloop) in the subband $n=2,1\uparrow$,
$\nu=4+\frac{1}{5(p-1)}$, $\nu=4+\frac{l}{5 l(p-2)\pm 1}$, respectively (inclusion of subband holes resolves itself to the substitution of $4+$ by $5-$ in both above formulae). Similarly as previously, the limit $l\rightarrow \infty$ gives the Hall metal hierarchy. 
The difference in comparison to the previous LL consists here also in the presence of four (instead two) satellite FQHE(singleloop) states symmetrically located around the central paired state. In the subband $n=2,1,\uparrow$ the satellite states occur at $\nu= \frac{21}{5}, \frac{22}{5}, \frac{23}{5}, \frac{24}{5}$ and the central paired state at $\nu= \frac{9}{2}$. This hierarchy is repeated in all 4 subbands of the first  LL. 

\begin{figure}[ht!]
\centering
\scalebox{0.6}{\includegraphics{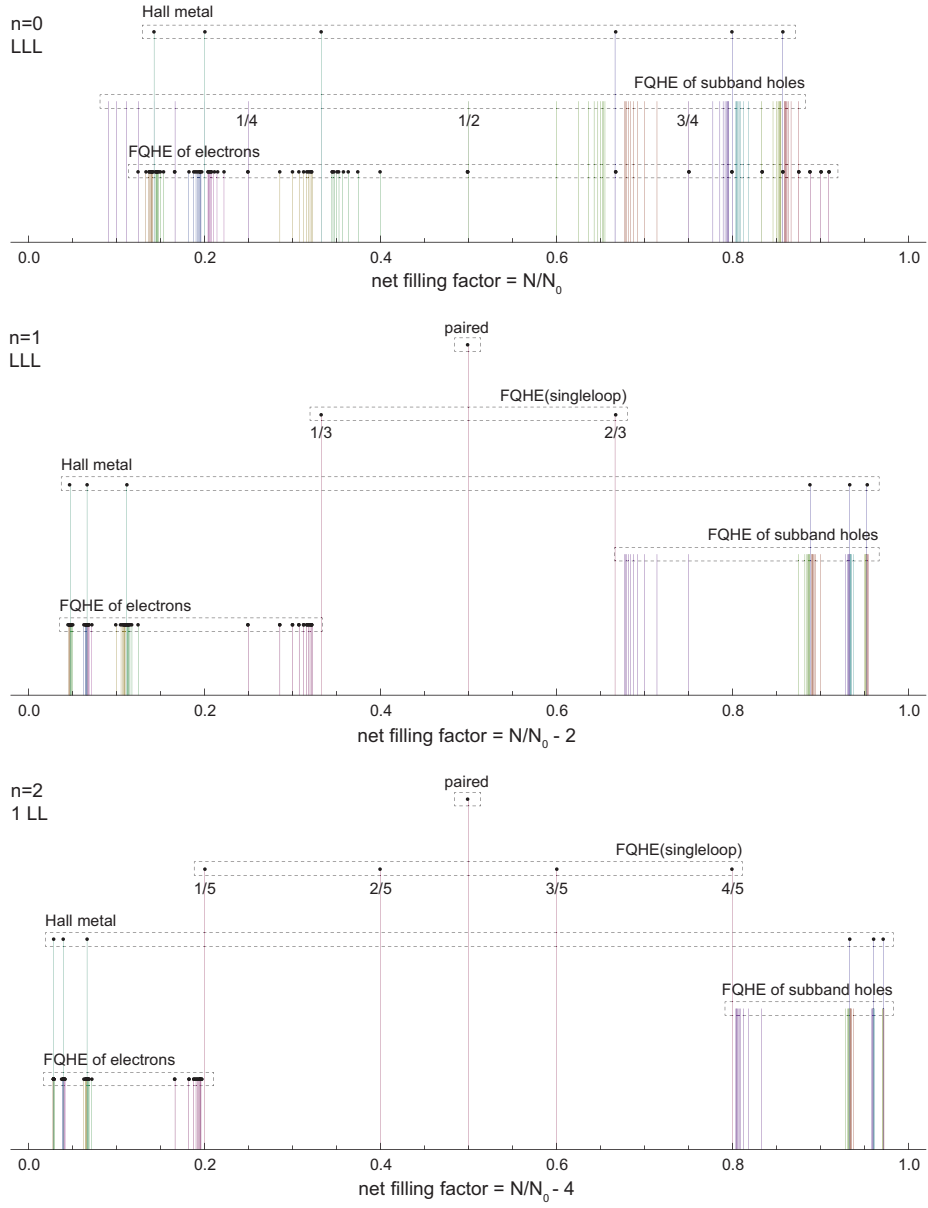}}
\caption{\label{figg5} Evolution of fractional filling hierarchy in two first LLs of bilayer graphene; for the LLL two subbands with $n=0$ and $n=1$ are  illustrated. Different types of ordering are indicated with spikes of various height. Series for FQHE (ordinary---multiloop), FQHE(singleloop), Hall metal and paired state are displayed acc. to the hierarchy described in Tab. \ref{tab2} with $q=3-9$, $l=1-10$; only a few selected ratios from these series  are explicitly  written out. }
\end{figure}

The evolution of the fractional filling hierarchy of subsequent LLs is illustrated in Fig.\ref{figg4} and Fig. \ref{figg5}, for monolayer and bilayer graphene, correspondingly and is summarized in Tab. \ref{tab1} and Tab. \ref{tab2}.

For bilayer graphene the degeneration of  $n=0$ and $n=1$ states results in 8-fold degeneration of the LLL doubling 4-fold spin-valley degeneration. The degeneration is not exact and with rising magnetic field amplitude both the Zeeman splitting and the  valley splitting grows. Stress, deformation and structure imperfections also cause the increase of the valley splitting. Inclusion of the interaction plays a similar role. Coulomb interaction causes mixing of $n=0,1$ states lifting their degeneration. Especially interesting is  such a degeneration lifting  which admits inverted order of fillings of LLL subbands with distinct $n$. The inversion of orders $n=0,1$ to $n=1,0$ affects the filling ratio hierarchy. Assuming that the LLL subbands with $n=1$ is earlier filled than the $n=0$ subband, we get the following hierarchy for  the first subband $n=1,2\uparrow$: multilooped orbits for $\nu=\frac{l}{3l(p-1)\pm 1}$, $\nu=1-\frac{l}{3l(p-1)\pm 1}$, siglelooped orbits for $\nu=\frac{1}{3},\frac{2}{3}$ and paired state for $\nu=\frac{1}{2}$. Assuming the next subband, $n=0,2,\uparrow$, we get the hierarchy of fillings for this subband in the form: multilooped orbits for $\nu=1+\frac{l}{l(p-1)\pm 1}$,
$\nu=2-\frac{l}{l(p-1)\pm 1}$ and no singlelooped orbits. The comparison of reverted orderings of two first LLL subbands is summarized in Tab. \ref{tab3}.    

One can consider also the situation in the LLL of bilayer graphene, when the degeneration of $n=0,1$ states is lifted is such a way that both levels cross at certain filling factor $\nu^*<1$ (cf. \cite{papec} where mixing between $n=0,1$ states is numerically analyzed for small models on torus or sphere). Let us assume for a model, that  first is filled the $n=1$ subband ($n=1,2\uparrow$) up to $\nu^*$. At this filling the subband $n=1,2\uparrow$ crosses with the subband $n=0,2\uparrow$ and the latter is filled for $1+\nu^*>\nu>\nu^*$. The related hierarchy of fractional fillngs looks like an ordinary filling of the subband $n=1,2\uparrow$, however, with an insertion of $n=0,2\uparrow$. Depending on the value of $\nu^*$ the various patterns are achievable by simple combination of hierarchy patterns listed in Tab. \ref{tab3} (including also inverted ordering of $n=1$ and $n=0$ subbands).

\section{Comparison with experiment}   

\begin{figure}[h]
\centering
\scalebox{1.0}{\includegraphics{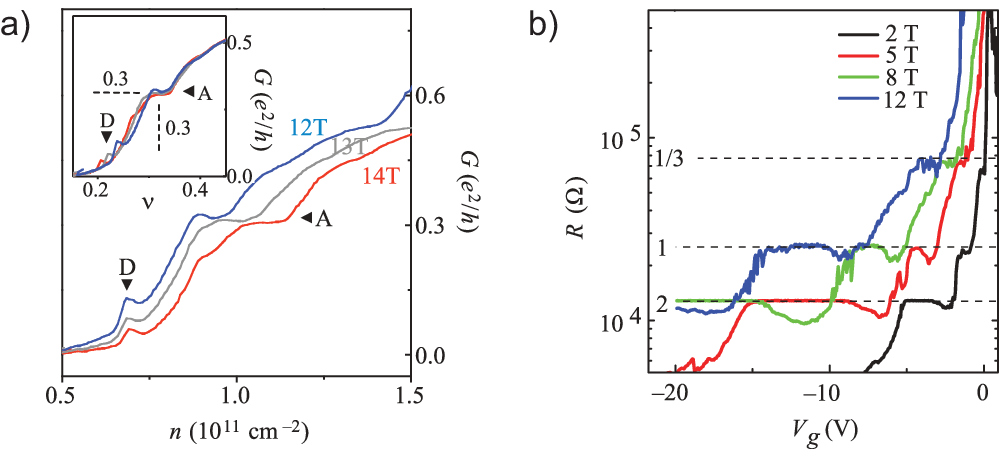}}
\caption{\label{grafenrys3}
 a) FQHE observation in suspended graphene for the filling 0.3 ($1/3$) in a field of 12-14 T with the concentration of $10^{11}$cm$^{-2}$ and the mobility of 250000 cm$^2$V$^{-1}$s$^{-1}$, b) FQHE singularities in suspended graphene for the filling $\frac{1}{3}$ in a field of 2-12 T with the concentration of $10^{10}$cm$^{-2}$ and the mobility of 200000 cm$^2$V$^{-1}$s$^{-1}$ (after \cite{fqhe1,fqhe2})}
\end{figure}

Searching states related to FQHE in the case of Hall  measurements in graphene is particularly challenging since the  different 'relativistic' structure of LLs more complicated in comparison to the conventional semiconductor 2DEG Hall physics. Moreover the filling factor can be changed in graphene both by the external magnetic field and by the particle concentration via shifting of the Fermi level near the Dirac point by application of the lateral voltage. Due to spin-valley degeneration and the Berry phase contribution related to chiral valley pseudospin, the IQHE is observed in graphene for fillings $\nu=4(n+\frac{1}{2})=2,6,10,14,..$ for particles from the conduction band and for the same but negative factors for holes from the valence band.  Despite using very strong magnetic fields (up to 45 T), FQHE was, however, not detected in graphene samples deposited on a substrate of $SiO_2$ \cite{dur1}. Instead, at so strong magnetic fields  the emergence of additional plateaus  of IQHE has been observed  for the fillings $\nu = 0,\pm 1, \pm 4$, indicating the elimination of spin-valley degeneration, as a result of increasing mass of Dirac fermions \cite{dur1}. Only after mastering the novel technology of the so-called suspended ultrasmall graphene scrapings with extreme purity and high mobility of carriers above 200000 cm$^2$V$^{-1}$s$^{-1}$ (high mobility is necessary to observe FQHE also in the case of conventional semiconductor 2D hetero-structures, which may be related to multi-looped quasi-classical cyclotron movement of wave packets in the case of multi-looped braids associated with FQHE \cite{jetph}; note that in  semiconductor 2D heterostructures carrier mobility reaches  even higher values of millions cm$^2$V$^{-1}$s$^{-1}$ \cite{pfeiffer2003}),  it was possible to observe FQHE in graphene at net fillings $\nu=1/3$ and $-1/3$ (the latter for holes, at the opposite polarization of  the gate voltage, which determines position of the Fermi level, either in the conduction band, or in the valence band) \cite{fqhe1,fqhe2}. Both these papers report the observation of FQHE in graphene for medium strong magnetic fields. The paper \cite{fqhe1}, in a field of 14 T, for electron concentration of $10^{11}$/cm$^2$ and  the paper \cite{fqhe2}, in a field of 2 T, but for a concentration level smaller by one order  of magnitude ($10^{10}$cm$^{-2}$ and the mobility of 200000 cm$^2$V$^{-1}$s$^{-1}$) as shown in Fig. \ref{grafenrys3}.

FQHE in suspended graphene is observed at the temperatures around 10 K \cite{fqhe3}, and even higher (up to 20 K) \cite{fqhe6}, which seems to be related with the stronger electric interaction in view of lack, in the case of suspended samples, of a dielectric substrate (with the dielectric constant in case of $SiO_2$, $\sim 3.9$) and, on the other hand, with very high cyclotron energy in graphene (i.e., large energy gap between incompressible states).

\begin{figure}[h]
\centering
\scalebox{1.0}{\includegraphics{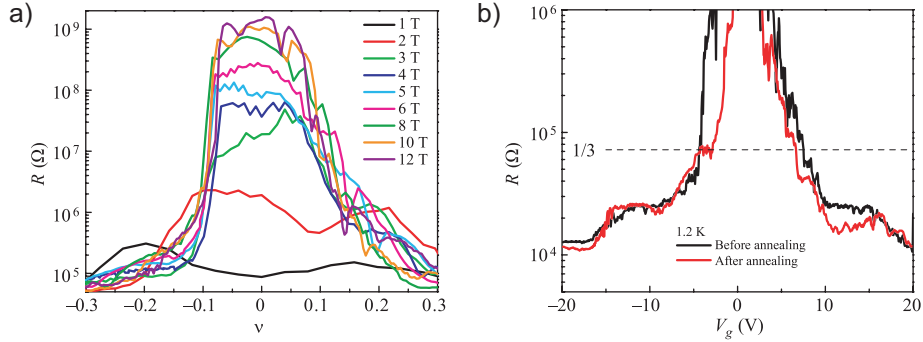}}
\caption{\label{grafenrys4}a) The emergence of an insulator state accompanying the increase in the strength of a magnetic field around the Dirac point, b) competition between FQHE and the insulator state for the filling $-1/3$: annealing removes pollution---enhances mobility and provides conditions for the emergence of {\it plateau} for FQHE (after \cite{fqhe2})}
\end{figure}

In the papers \cite{gr5,clure} it has also been demonstrated the competition between the FQHE state with the insulator state near the Dirac point, corresponding to a rapidly decreasing concentration---Fig. \ref{grafenrys4}. 

From the perspective of cyclotron groups, experimental results on FQHE in graphene \cite{fqhe1,fqhe2,fqhe3,fqhe6,fqhe4} seem to be compliant with the expectations of the braid description.
In the case of graphene, the specific band structure  with conical Dirac bands leads to simultaneous participation (in  Dirac point) of both bands---of holes and of electrons, which combined with the massless character of Dirac fermions manifests itself through an anomalous 'relativistic' IQHE  \cite{gr4,gr5,dur1}. Controlling lateral gate voltage (within the range ca. $5-60$ V \cite{fqhe1,bil,dean2011}) allows regulating the density of carriers at a constant magnetic field. One should therefore expect  that at relatively  small densities of carriers (electrons, or symmetrical holes at reverse voltage polarization), the  cyclotron orbits will be too short to prevent braid exchanges of particles at a sufficiently strong magnetic field---although weaker for smaller concentrations---and experimental observations exactly support that \cite{fqhe1,fqhe2}. For low concentration, while closing on the Dirac point, one may expect that too strong fields would exceed the stability threshold of the FQHE state in competition with the Wigner crystal (assuming a similar character of this competition in the case of massless Dirac fermions in reference to conventional semiconductor 2D structures) and that corresponds to the emergence of the insulating state near the Dirac point in a sufficiently large  magnetic field \cite{fqhe5}. In the case of the hexagonal structure of graphene, electron (or hole) Wigner crystallization \cite{wigno} may exhibit interference between the triangular crystal sublattices, and including of the resonance (hopping) between these two sublattices may cause blurring the sharp transition to the insulator state, which seems compliant with observations (Fig. \ref{grafenrys4}).

The recent progress in the experiment allows also for observation of FQHE in graphene on the crystal substrate of boron nitride  ($BN$)  in large magnetic fields of order of 40 T (remarkably, FQHE features were noticed  in this case up to $\nu=4$) \cite{dean2011}.

The mobility of carriers in graphene is lower than in traditional 2DEG, but  taking into account that the carrier concentration in graphene can be lower in comparison to semiconductor heterostructure \cite{pfeiffer2003,bolotin2008}, the corresponding mean free path in both cases well exceeds the sample dimension (of $\mu$m order, as the mobility is proportional to the concentration and to the mean free path of carriers).

 Energy gaps protecting incompressible FQHE states are in graphene larger than in traditional semiconductor materials, reaching an order of 16 K (at $\nu=\frac{4}{3}$ and $B=35$ T), which is referred to Dirac massless character of carriers. In conventional semiconductor heterostractures the corresponding gaps are much lower and the observed FQHE stability with temperature is much more fragile.

\begin{figure}[h]
\centering
\scalebox{1.0}{\includegraphics{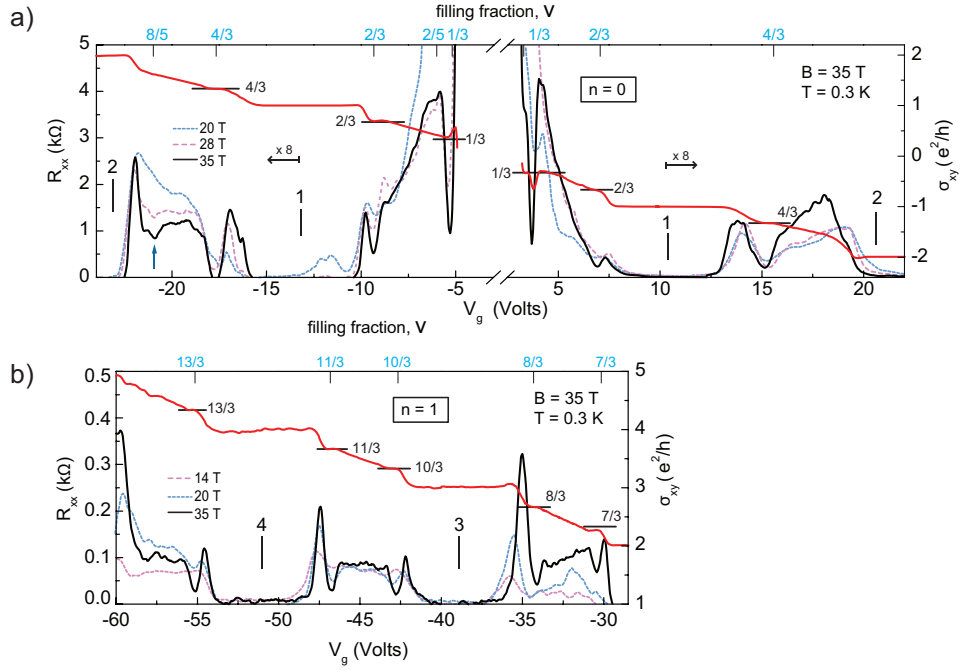}}
\caption{\label{grafenrys555} Fractional quantum Hall effect for graphene on BN, a,b. Magnetoresistance (left axis) and Hall conductivity (right axis) in the $n=0$ and $n=1$ Landau levels at
$B=35$ T and temperature $ \sim 0.3$ K (after \cite{dean2011}). All filling ratios indicated in blue agree with the hierarchy given in Tab. \ref{tab1}.}
\end{figure}
\begin{figure}[h]
\centering
\scalebox{1.0}{\includegraphics{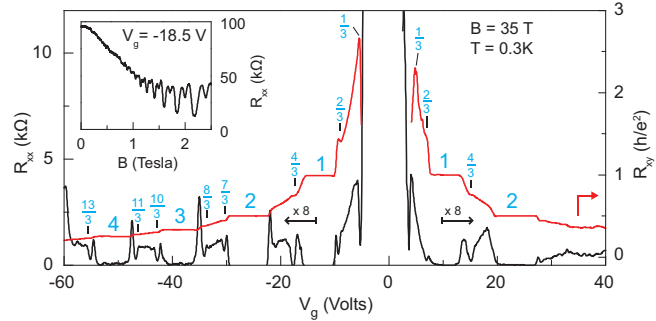}}
\caption{\label{grafenrys555-2} 
Magnetoresistance (left axis) and Hall resistance for graphene on BN (right
axis) versus gate voltage acquired at B = 35 T. Inset shows
SdH oscillations at Vg = -18.5 V (after \cite{dean2011}). All filling ratios indicated in the figure (in blue) agree with the hierarchy as given in Tab. \ref{tab1}. }
\end{figure}

The recent development in experiments with monolayer graphene on BN substrate \cite{dean2011,amet} and with suspended small sheets \cite{feldman,feldman2} allowed for observation of more and more Hall features at fractional fillings of subsequent subbands of two first LLs. While the sequence of fillings in the lowest subband of the LLL fits well to CF predictions (including CFs with two and four flux quanta attached), an explanation of the filling structure of next subbands strongly deviates from this simple picture. The pattern of filling rates repeated in the subbands  of the first LL has apparently nothing in common with the CF concept \cite{feldman, feldman2, amet}. This phenomenon is referred in these papers to the various scenarios of breaking of the approximate SU(4) spin-valley symmetry in graphene. Because of the smallness of the Zeeman splitting  $E_Z$   in comparison to the Coulomb energy in graphene, $E_Z/E_C\sim 0.01\varepsilon$, similarly as  of lattice scale in comparison to the magnetic length, $a/l_B\sim 0.06$ (where $l_B=\sqrt{\frac{h}{eB}}$ and $\varepsilon $ is the dielectric susceptibility $\sim 3.2$) \cite{dean2011},  the subbands which differ with spin and valley-pseudospin orientation are closely located and can be regarded as approximately degenerated. This SU(4) spin-valley symmetry can be next broken by various factors and one can search arguments for unusual filling ratios hierarchy in related symmetry breaking and phase-like transitions. 
Despite of many related ideas no fully consistent picture is attained in this way as of yet. 

If one compares the experimentally observed fractions for characteristic FQHE features in longitudinal and Hall conductivities measured on variety of samples of graphene with the  pattern of fractional hierarchy for two lowest LL, as  illustrated in Fig. \ref{figg4}, one  notices the  coincidence of this hierarchy with the measured data. All fractions found experimentally can be reproduced by this hierarchy (cf. Tab. \ref{tab1}).  From this comparison it is  visible why the CFs are efficient  only in the LLL. This is linked with the fact that exclusively in the LLL cyclotron orbits are always shorter than the interparticle separation and additional loops are necessary. These loops can be modeled by fictitious field flux quanta attached to CFs. Though the analogy to additional loops is not exact, it  allows to get  at least  the similar main line of the filling hierarchy in the LLL as that one given by the commensurability condition. The usefulness of the CF model dramatically breaks down, however, in  higher LLs because starting from  the first LL the multilooped commensurability is needed only close to the subband borders, whereas the central regions of all subbands of the first LL are occupied by doublets of filling factors, ($\frac{7}{3},\frac{8}{3}$), ($\frac{10}{3},\frac{11}{3}$), ($\frac{13}{3},\frac{14}{3}$), ($\frac{16}{3},\frac{17}{3}$), corresponding to singlelooped commensurability condition, not allowing for CF modeling, but  visible in experiments as FQHE(singleloop) \cite{feldman, feldman2, amet, dean2011}. The number of centrally located filling rates for FQHE(singleloop) grows next with the LL number as $2n$. The repeating doublet  of filling ratios for $n=1$ is noticeably e.g., in Fig. \ref{grafenrys555-2} and in more accurate measurements in suspended samples \cite{feldman,feldman2} besides of those on the BN substrate \cite{dean2011}.

The hierarchy induced by the commensurability condition reproduces the position of other observed features in two lowest LLs of the monolayer graphene. The elongated plateaus at border of subbands with IQHE seem to overwhelm also minima related by closely to border located FQHE rates being in this way out of the experimental resolution.

\begin{figure}[h]
\centering
\scalebox{0.4}{\includegraphics{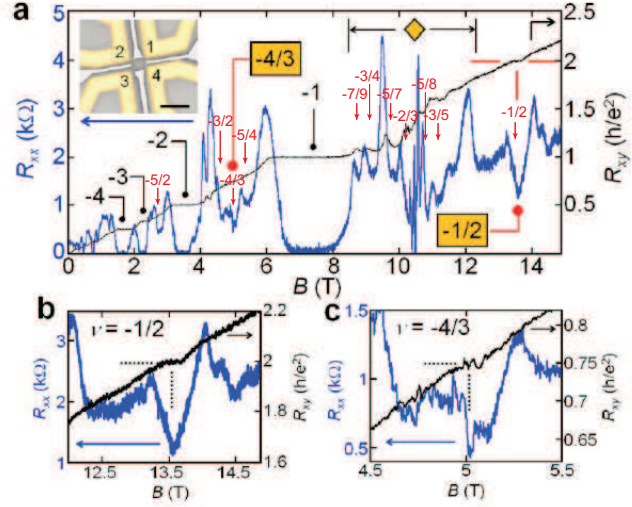}}
\caption{\label{figg6} Observation of FQHE at T = 0.25 K in bilayer suspended graphene. a) magneto-resistance $R_{xx}$ (blue curve) and $R_{xy}$ (black curve) at the lateral voltage $ -27$ V (the inset shows an image of the actual device covered with a resist mask).  Zoom-in on the data shown in a) to better illustrate the features associated with the $-1/2$ and $-4/3$ FQHE states, respectively; the features in the range
marked by the yellow diamond are affected by remnant disorder (after \cite{bil}). In red is added fitting with the hierarchy given in  Tab. \ref{tab2}.}
\end{figure}

The most convincing evidence supporting the correctness of the commensurability condition is, however, the coincidence of the related predictions with the experimental observations in the bilayer graphene. The pronounced feature of these observations is the occurrence  of  primary fractional features with even denominators also in the lowest subband of the LLL in  bilayer graphene, oppositely to monolayer one. \cite{bil}. The commensurability condition for bilayer graphene reproduces perfectly the observed experimentally hierarchy---cf. Fig. \ref{figg6}, as is  illustrated in Fig. \ref{figg5} and summarized  in Tab. \ref{tab2}.

Note finally that the FQHE hierarchy in bilayer systems with the characteristic even denominators holds also for bilayer 2DEG conventional Hall setups and indeed the $\nu=\frac{1}{2}$ state has been discovered there \cite{2degbil,2degbill}.

\section{Conclusion}       
The condition for commensurability of cyclotron orbits building cyclotron braid subgroups with interparticle spacing in uniform 2D charged systems is formulated in order to verify possibility of arrangement  of correlated multiparticle states in planar quantum Hall systems.  
Using this commensurability condition based on the braid group approach to statistics of many particle systems the hierarchy of fractional fillings for LLs in graphene was determined. The FQHE evolution with growing number of LL has been described. The new opportunities for commensurability in higher LLs were established leading to different that ordinary FQHE(multiloop) correlated states at fractional fillings of LLs starting from the first one have been identified (they are referred to FQHE(singleloop) as the related correlations are described by singlelooped braids). Both the monolayer and bilayer graphene were considered and the essential difference of related hierarchy structures has been demonstrated and described. The even denominator main line of the fractional filling hierarchy in bilayer graphene is found in agreement with the experimental observations. The presented hierarchy for the monolayer and bilayer graphene found the  confirmation in available experimental data for graphene on BN substrate as well as for suspended samples including bilayer suspended sheets for fillings of related  spin-valley subbands from the zeroth and the first Landau levels.

\begin{acknowledgments}
The support from the NCN Project UMO-2011/02/A/ST3/00116 is acknowledged.
\end{acknowledgments}

\end{document}